\documentstyle[prl,aps,epsfig,multicol,amssymb,amsfonts]{revtex}
\tighten
\renewcommand{\narrowtext} 
{\begin{multicols}{2}\global\columnwidth20.5pc} 
\renewcommand{\widetext}
{\end{multicols}\global\columnwidth42.5pc} 
\newcommand \be {\begin{equation}}
\newcommand \ee {\end{equation}}
\newcommand \bea {\begin{eqnarray}}
\newcommand \eea {\end{eqnarray}}

\begin{document} 
\draft 
\title{Quasiclassical negative magnetoresistance of a 2D electron gas:
interplay of strong scatterers and smooth disorder} 
\author{A.~D.~Mirlin,$^{1,2,*}$  D.~G.~Polyakov,$^{1,\dagger}$
F.~Evers,$^1$ and P.~W\"olfle$^{1,2}$}  
\address{$^1$Institut
f\"ur Nanotechnologie, Forschungszentrum Karlsruhe, 76021 Karlsruhe,
Germany}
\address{$^2$Institut f\"ur Theorie der Kondensierten Materie,
Universit\"at Karlsruhe, 76128 Karlsruhe, Germany}
\date{April 5, 2001}
\maketitle
\begin{abstract}

We study the quasiclassical magnetotransport of
non-interacting fermions in two dimensions
moving in a random array of strong scatterers (antidots, impurities or
defects) on the background of 
a smooth random potential. We demonstrate that the combination of 
the two types of disorder induces a novel mechanism leading to
a strong negative magnetoresistance, followed by the saturation of the
magnetoresistivity $\rho_{xx}(B)$ at a
value determined solely by the smooth disorder. Experimental relevance
to the transport in semiconductor heterostructures is discussed.

\end{abstract}

\narrowtext

Magnetotransport in a two-dimensional electron gas (2DEG) has been the
subject  of intensive research during the last two decades. This
interest has been motivated by the progress in preparing
high-quality semiconductor heterostructures, opening up new areas in
both fundamental physics and applications, see
\cite{transport-review} for a review. Within the quasiclassical
approach (valid for not too strong magnetic fields $B$), impurity
scattering is commonly described by a collision integral in the
Boltzmann equation. This leads, for an isotropic system, to the 
$B$-independent Drude value of the longitudinal resistivity, 
$\rho_{xx}(B)=\rho_0\equiv m/e^2n_e\tau$, where $n_e$ is
the carrier density, $m$ the effective mass, and $\tau$ the transport
scattering time.

It has become clear, however, that this description is not always
valid. In particular, in the case of smooth disorder memory effects
induce a strong positive magnetoresistance (MR) \cite{m99} followed
by an 
exponential fall-off of $\rho_{xx}(B)$ due to adiabatic localization
of drifting electrons \cite{fogler97,m98}.  To our knowledge, these
effects have not been experimentally observed in the  electron
transport in low magnetic fields, since the  Shubnikov-de Haas
oscillations  develop at lower magnetic fields.  
On the other hand, the memory effects do
show up in the composite fermion transport, explaining the peculiar
shape of the MR around half-filling of the lowest
Landau level \cite{m98,m99}.

In the present paper we study the quasiclassical MR of a 2DEG
moving in a random array of rare strong scatterers (modeled by
hard disks) and subject additionally to a smooth
random potential. Apart from the purely theoretical interest, our work
has been motivated by two types of experimental realizations of this
problem. The first one is random antidot arrays. Experiments on
this kind of structures
\cite{gusev94,tsukagoshi95,luetjering96,nachtwei97} show a strong
negative MR which has not been analyzed theoretically. 
Less obviously, our model is relevant to transport in
the unstructured high-mobility 2DEG. To clarify this point, we recall
that in order to increase the 2DEG mobility the donors in currently
fabricated heterostructures are separated by a  large
distance $d\gg k_F^{-1}$ (with $k_F$ the Fermi wave number) 
from the 2DEG plane. It is usually assumed that these
remote donors constitute the main source of disorder, inducing
small-angle scattering of electrons. It is known, however, that in
samples with a wide spacer ($d\gtrsim 70\, {\rm nm}$) large-angle
scattering on residual impurities \cite{coleridge91,saku96,umansky97}
and interface roughness \cite{saku96} becomes important,
limiting the mobility with further increasing width of the
spacer. In particular, Ref.~\cite{saku96} concludes that about 50\% of
the resistivity is determined by such scattering processes, while
Ref.~\cite{umansky97} finds that for samples with very high mobility
this value is as high as $90\%$.

We thus consider the following two-component model of disorder: (i)
randomly distributed hard-core scatterers (which we will term ``antidots''
or ``impurities'' below) with density $n_S$ and
radius $a$ (where $n_S^{-1/2}\gg a \gg k_F^{-1}$), 
and (ii) smooth random potential (correlation radius $d$,
momentum relaxation rate $\tau_L^{-1}$, transport mean free path
$l_L=v_F\tau_L$). The mean free path for the scattering on antidots is 
$l_S^{(0)}\equiv v_F\tau_S^{(0)}=1/2n_Sa$, while the corresponding
transport mean 
free path may be somewhat different, $l_S\equiv v_F\tau_S=\gamma
l_S^{(0)}$ with $\gamma\sim 1$ ($\gamma=3/4$ in the model of
specularly reflecting disks). We will set $\gamma=1$ for qualitative
estimates.  
We will further assume that $\tau_L\gg\tau_S$, so that the zero-$B$
resistivity $\rho_0$ is determined by the hard scatterers,
$\tau^{-1}=\tau_L^{-1}+\tau_S^{-1} \simeq
\tau_S^{-1}$. Finally, we will assume that the motion in the smooth
disorder is not adiabatic, {\it i.e.} has the form of the guiding center
diffusion (rather than drift). The condition for this is
$\delta\gg d$, where $\bbox{\delta}$ is the guiding center shift after
one cyclotron revolution [see Eq.~(\ref{e4})]. 
For currently fabricated samples
this assumption is usually satisfied in the whole range of
applicability of the quasiclassical theory. 

We start the analysis of the problem by recalling the results
\cite{baskin78,bobylev95} for the case when only hard scatterers are
present ($\tau_L=\infty$), known as the Lorentz gas.  In this limit the
resistivity $\rho_{xx}(B)$ reads 
\be
\label{e3}
\rho_{xx}(B)/\rho_0=(1-e^{-2\pi/\omega_c\tau_S^{(0)}})
{\cal F}(\omega_c\tau_S), 
\ee
where $\omega_c=eB/mc$ is the cyclotron frequency and ${\cal F}(x)$ is
a function of order unity with the asymptotics ${\cal F}(x\ll 1)=1$
and ${\cal F}(x\gg 1)=\gamma$. The first factor in Eq.~(\ref{e3}) is
nothing else but the fraction of particles moving in rosette-like
trajectories around the impurities and hitting a new impurity with the mean
free time $\tau_S^{(0)}$. The rest of the particles do not hit scatterers at
all. In the sequel, we will only consider classically strong
magnetic fields, $\omega_c\gg\omega_\tau\equiv 2\pi/\tau_S^{(0)}$, where
the resistivity (\ref{e3}) shows a $1/B$ fall-off.
Equation (\ref{e3}) is valid below the percolation threshold, 
$\omega_c<\omega_{\rm perc}=1.67 v_Fn_S^{1/2}$ (note that 
$\omega_{\rm perc}\tau_S\sim n_S^{-1/2}a^{-1}\gg 1$). For larger magnetic
fields, $\omega_c\ge \omega_{\rm perc}$, the resistivity is exactly
zero, since the rosette-like families of cyclotron orbits associated
with each impurity fail to form an infinite cluster.  

Clearly, adding the long-range disorder will increase the diffusion
constant $D_{xx}$ and thus, in the limit $\omega_c\tau_S\gg 1$,  the
longitudinal resistivity, by   
setting free those particles which are localized in cyclotron
orbits not hitting impurities. We will be
interested in the case of a sufficiently strong smooth disorder
modifying the result (\ref{e3}) in an essential way. 
Specifically, we will see below that new physics
emerges in the regime  $\delta\gg a$, where 
\be
\label{e4}
\delta^2\equiv\langle\bbox{\delta}^2\rangle=
4\pi l_L^2/(\omega_c\tau_L)^3.
\ee

Let us first outline this new physics on a qualitative level. 
Naively, one could think that for $\delta\gg a$ the resistivity should
take its Drude value. Indeed, let us associate with the particle
trajectory a strip of width $2a$ surrounding it. The particle will hit an
impurity if the center of the latter 
is located within this strip. Clearly, in one
cyclotron revolution the particle ``explores'' in this way the area
$2v_Fa\times 2\pi/\omega_c$, the same as it would explore in the
same time for $B=0$. For $\delta\gg a$ the area explored in the second
revolution will overlap only weakly with that explored in
the first one, so that one could think that the exploration rate
is essentially the same as for $B=0$, leading to the mean
time $\simeq\tau_S^{(0)}$ between the collisions, and thus to
$\rho_{xx}\simeq\rho_0$. This consideration is, however, incorrect, since
it neglects memory effects. Specifically, there is a probability
$P_1\sim a/\delta$ that the strip after the first revolution covers again
the starting point (or, in other words, there is typically a small
relative overlap $\sim a/\delta$). In view of the diffusive
dynamics of the guiding center, the return probability decreases rather
slowly with the number of revolutions, $P_n=P_1/\sqrt{n}$. Therefore,
the total return probability $P=\sum_{n=1}^N P_n$ is determined by the
upper cut-off $N$, so that the memory effect is much stronger than 
one might at first sight expect, $P\sim (a/\delta)N^{1/2}\gg a/\delta$.

When the memory
effect leads only to a small correction to the Drude value (the
corresponding condition will be specified below),  the
upper cut-off is given by the number of cyclotron revolutions it takes
for the particle to hit the next impurity, 
$N=\omega_c\tau_S^{(0)}/2\pi$, so that the total return
probability is $P\sim (a/\delta)(\omega_c\tau_S)^{1/2}$. This
determines the fraction of the area explored twice, implying an effective
reduction of the exploration rate and thus a negative correction to
the resistivity,
\be
\label{e5}
\Delta\rho_{xx}/\rho_0 \sim - 
(a/\delta)(\omega_c\tau_S)^{1/2} \sim -(\omega_c/\omega_0)^2,
\ee
where 
$\omega_0\sim v_F(a^2l_Sl_L)^{-1/4}\sim 
\omega_{\rm perc}(l_S/l_L)^{1/4}  \ll \omega_{\rm perc}$.

We turn now to a more rigorous and quantitative derivation. 
Generalizing the formalism of \cite{m99} to the present case, we start
from the following Liouville-Boltzmann equation for
the distribution function $g({\bf r},\phi)$ of electrons on the
Fermi surface,
\bea
&& (L_0+\delta L) g({\bf r},\phi)=\cos(\phi-\phi_E), 
\label{e1} \\
&& L_0 = v_F{\bf n\nabla}+ \omega_c{\partial \over\partial
\phi}-{1\over \tau_L}{\partial^2\over\partial\phi^2},
\label{e2}\\
&&\delta L = -\sum_i I_{{\bf R}_i}, 
\label{e2a}
\eea
where ${\bf n}=(\cos\phi,\sin\phi)$ is the unit vector determining the
direction of velocity, ${\bf v}=v_F{\bf n}$, and
$\phi_E$ is the polar angle of the electric field.
The operator $L_0$ describes the
motion in the smooth random potential, 
while $\delta L$ corresponds to the scattering on
antidots with (random) positions ${\bf R}_i$. The explicit form of the
collision operator $I_{\bf R}$ (for a hard-wall scatterer) can be
found in \cite{vanleeuwen}, but we will not need it. We will only use
the following properties of $I_{\bf R}$: (i) it is non-zero only
within the distance $a$ from the impurity location ${\bf R}$, and (ii)  
its Fourier-transform $\tilde{I}_{\bf q}$ satisfies 
$\tilde{I}_0{\bf n}=(-1/n_S\tau_S){\bf n}$. 

Expanding in $\delta L$, averaging over the positions ${\bf R}_i$ of
scatterers, and resumming the series, one finds the Green's function
of the Liouville operator, $\langle(L_0+\delta L)^{-1}\rangle =
(L_0+M)^{-1}$, where $M$ is the ``self-energy'' operator. The
resistivity is expressed in terms of $M$ as follows \cite{m99},
\be
\label{e8}
\rho_{xx}=(m/ e^2 n_e)(\tau_L^{-1}+M_{xx}), 
\ee
where $M_{xx}=\int(d\phi/\pi)\cos\phi\, M\cos\phi$. 
One may use a diagrammatic technique analogous to that developed for
the Lorentz gas \cite{vanleeuwen,hauge74} to calculate 
$M$. The leading term corresponds to a single
scattering process, yielding $M^{(1)}=-n_S\tilde{I}_0$ and thus
the Drude contribution, $M_{xx}^{(1)}=\tau_S^{-1}$, to the resistivity
(\ref{e8}). 
The next-order contribution representing the correction induced by 
the return process is obtained as
\be
\label{e10}
M_{xx}^{(2)}=-n_S\int {d\phi\over\pi}\cos\phi\, 
I_{\bf R}D^{(1)} I_{\bf R}\cos\phi,
\ee
where $D^{(1)}=(L_0+M^{(1)})^{-1}$ is 
the electron propagator with the
leading-order self-energy included.

%
Since $\delta\gg a$, the propagator 
$D^{(1)}({\bf r}-{\bf r'},\phi,\phi')$ describing propagation from the
point ${\bf r'},\phi'$  to the point ${\bf r},\phi$
can be replaced in (\ref{e10}) by
$D^{(1)}(0,\phi,\phi')$. Furthermore, we note that once the particle hits
an impurity, its guiding center is shifted by an amount of order of the
cyclotron radius $R_c$. As a result, the contribution of such
trajectories to $D^{(1)}$ can be neglected and only non-colliding
orbits should be taken into account. 
Since the motion without collisions is
limited by the times $\sim\tau_S\ll\tau_L$, the particle will return
with almost the same direction of velocity, {\it i.e.}
$D^{(1)}(0,\phi,\phi')$ will be peaked at $\phi\approx\phi'$. We can
thus approximate $D^{(1)}(0,\phi,\phi')$ by
$D_0^{(1)}\delta(\phi-\phi')$, where $D_0^{(1)}=\int d\phi
D^{(1)}(0,\phi,\phi')$. This quantity is easily found to be
\be
\label{e11}
D_0^{(1)}=\sum_{n=1}^\infty{e^{-2\pi n/\omega_c\tau_S^{(0)}}  
\over(\pi n)^{1/2}v_F\delta} \simeq 
{(\omega_c\tau_S^{(0)})^{1/2}\over  (2\pi)^{1/2} v_F\delta}.
\ee
Substituting this in (\ref{e10}) and using that 
$\tilde{I}_0{\bf n}=-(1/n_S\tau_S){\bf n}$, we find
$M_{xx}^{(2)}=-(1/ n_S\tau_S^2)D_0^{(1)}$.
This implies, according to (\ref{e8}), the negative MR,
\be
\label{e13}
\Delta\rho_{xx}/\rho_0= M_{xx}^{(2)}/\tau_S^{-1}
=-\omega_c^2/\omega_0^2, \qquad \omega_c\ll\omega_0,
\ee
where 
\be
\label{e14}
\omega_0= (2\pi n_S)^{1/2}v_F(2\gamma l_S/l_L)^{1/4},
\ee
in agreement with the above qualitative considerations.

%
%

This derivation is valid as long as the correction remains small,
i.e. for $\omega_c\ll \omega_0$. For stronger magnetic fields,
when the quantity $P$ defined above becomes large, $P\gg 1$, it
acquires the meaning of the number of returns. It should then be found
self-consistently. Specifically, the sum over $n$ is now cut off at
the time 
$\tau_S'\sim P\tau_S$, since multiple returns to the same area lead to
a corresponding increase of the time needed to hit a new
impurity. We thus get a self-consistency equation, 
$P\sim (a/\delta)(\omega_c P\tau_S)^{1/2}$, yielding a $1/B^4$ drop of
the resistivity,
\be
\label{e7}
\rho_{xx}/\rho_0 \sim \tau_S/\tau_S' \sim 1/P \sim
(\omega_0/\omega_c)^4, \qquad \omega_c\gg\omega_0.
\ee
  
It remains to analyze the conditions of validity of Eqs.~(\ref{e13}),
(\ref{e7}). 
First of all, we assumed that the particle finds a new scatterer by
exploring the new area in the course of the diffusive motion of the
guiding center. There exists, however, a competing mechanism, namely
that of the rosette-like motion around a scatterer, which determines
the transport in the pure Lorentz gas ($l_L=\infty$).  
Comparing (\ref{e7}) with the  Lorentz-gas result
$\rho_{xx}/\rho_0=2\pi\gamma^2/\omega_c\tau_S$ [see Eq.~(\ref{e3})], we find
that the two formulas match at $\delta\sim a$.  The corresponding
crossover frequency is  
$\omega_{\rm cross}=v_F(4\pi n_S^2 l_S^2l_L^{-1})^{1/3}$. Secondly,
we assumed $n_SR_c^2\gg 1$, or, equivalently, $\omega_c\ll\omega_{\rm
perc}$. It is easy to see that in the opposite limit, $n_SR_c^2\ll 1$,
the resistivity will be determined by the smooth disorder, with
scattering on antidots giving a small [$\sim(\omega_{\rm
perc}/\omega_c)^2$] correction only, so that 
$\rho_{xx}(B)$ will have a plateau with the
value $\rho_{xx}(\omega_c\gg\omega_{\rm perc})=m/e^2n_e\tau_L$.

\begin{figure}
\centerline{\epsfxsize=65mm\epsfbox{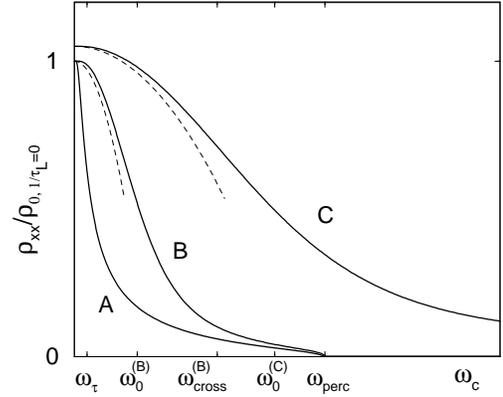}}
\caption{Magnetoresistivity at fixed $\tau_S$ and
different $\tau_L$: curve A -- Lorentz gas ($\tau_L=\infty$), curves B
and C with $\tau_L^{(B)}>\tau_L^{(C)}$ correspond to regimes B and C,
respectively (see text). The dotted lines denote the asymptotics
(\ref{e13}).}   
\label{fig2} 
\end{figure}

Comparing the characteristic frequencies,
$\omega_\tau$, $\omega_0$, $\omega_{\rm cross}$, and $\omega_{\rm
perc}$, we conclude that the following three situations can be
distinguished, depending on the strength of the smooth disorder
(Fig.~\ref{fig2}):

(A) $\omega_0\ll\omega_{\tau}$, or, equivalently, 
$l_L/l_S\gg (1/2\pi^2)(n_Sl_S^2)^2$. In this case the smooth disorder
hardly affects  the Lorentz-gas result (\ref{e3}).

(B) $\omega_\tau\ll\omega_0\ll \omega_{\rm cross}\ll \omega_{\rm
perc}$, or, equivalently, 
$2.7(n_Sl_S^2)^{1/2}\ll l_L/l_S\ll (1/2\pi^2)(n_Sl_S^2)^2$.
This is an intermediate situation; the resistivity drops first
according to Eqs.~(\ref{e13}), (\ref{e7}) and then crosses over at
$\omega_c\sim\omega_{\rm cross}$ to the Lorentz-gas behavior
(\ref{e3}).

(C) $\omega_\tau\ll\omega_0\ll \omega_{\rm perc}\ll 
\omega_{\rm cross}$, or, equivalently, 
$l_L/l_S\ll 2.7(n_Sl_S^2)^{1/2}$. In this case the Lorentz-gas
behavior (\ref{e3}) is completely destroyed, and the results
(\ref{e13}), (\ref{e7}) hold in the whole range of $\omega_c$ below 
$\omega_{\rm perc}$. 

For $\omega_c>\omega_{\rm perc}$ the resistivity
shows in all the cases a plateau, as explained above. On the side of
strong magnetic fields this plateau will be modified either by
entering into the adiabatic regime  at $\delta\sim d$, meaning
$\omega_c\sim \omega_{\rm ad}=v_F(4\pi/d^2l_L)^{1/3}$ (its
implications for 
the present problem will be considered elsewhere \cite{tobe}) or by
the development of Shubnikov-de Haas oscillations.

It is worth mentioning that while we used the condition
$\tau_L\gg\tau_S$ for the derivation of our main results,
a pronounced negative MR
will also be observed for $\tau_L\sim\tau_S$ (which seems to be
frequently the relevant situation for high-mobility structures
\cite{saku96}). In this case there is a crossover from 
$\rho_{xx}(0)=(m/e^2n_e)(\tau_L^{-1}+\tau_S^{-1})$ to 
$\rho_{xx}(\omega_c\gg\omega_{\rm perc})=m/e^2n_e\tau_L$ which takes
place around $\omega_c\sim\omega_{\rm perc}$ (note that
$\omega_0\sim\omega_{\rm perc}$ for  $\tau_L\sim\tau_S$).

\begin{figure}
\centerline{\epsfxsize=65mm\epsfbox{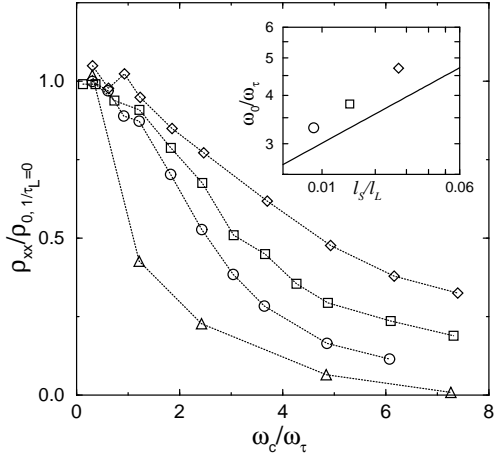}}
\caption{Magnetoresistivity at fixed $\tau_S$ and
different $\tau_L$ from numerical simulations; $\tau_L/\tau_S=\infty$
(Lorentz gas, $\triangle$), 111 ($\bigcirc$), 70 ($\square$), 37
($\Diamond$). Inset: $\omega_0$ determined from the fit to
Eq.~(\ref{e13}); the full line corresponds to the analytical result
(\ref{e14}).} 
\label{fig3} 
\end{figure}

We have performed numerical simulations of the MR by fixing parameters
of the Lorentz gas ($\omega_{\rm perc}/\omega_\tau=5.3$) and the
correlation length of the smooth disorder ($d/a\sim 2.5$) and changing
the strength of the latter. The results shown in
Fig.~\ref{fig3} are in good agreement with the analytical
predictions. It is seen that a very weak smooth disorder (giving
negligible contribution to $\rho_0$) affects crucially the MR.
Some deviations in the values of $\omega_0$ from
Eq.~(\ref{e14}) (see inset)
can be attributed to the fact that the conditions $\omega_\tau\ll
\omega_0\ll \omega_{\rm perc},\omega_{\rm ad}$ are fulfilled only
marginally in our simulations.

Finally, let us estimate the characteristic values of $B$
for existing experiments. Typical parameters in the experiments
\cite{luetjering96} on the antidot arrays were
$n_e= 5\times10^{11}\:{\rm cm}^{-2}$, 
$n_S= (0.6\:\mu{\rm m})^{-2}$,
$l_S=1.3\:\mu{\rm m}$, $l_L=16\:\mu{\rm m}$. This implies the following
values for the characteristic magnetic fields (we use the obvious
notations $B_\tau=(mc/e)\omega_\tau$ etc.):
$B_\tau\simeq 0.5\:{\rm T}$, 
$B_0\simeq B_{\rm perc}\simeq 0.3\:{\rm T}$. We see that the condition
of a dilute antidot array assumed above, $B_\tau\ll B_{\rm perc}$, is
not met. Clearly, in this situation 
the above analytical formulas are not valid quantitatively. On the
qualitative level, we can conclude that there should be a strong
fall-off of $\rho_{xx}$ around $B\approx 0.3\div 0.5\:{\rm T}$, in
agreement with experimental results \cite{luetjering96}. A similar
negative MR was observed in other experiments with
disordered antidot arrays \cite{gusev94,tsukagoshi95,nachtwei97}.

We turn now to unstructured high-mobility samples. Using the
parameters of \cite{umansky97}
($n_e=2\times 10^{11}\:{\rm cm}^{-2}$, mobility $\mu=10^7\:{\rm
cm}^2/{\rm V}\cdot{\rm s}$), we find the mean free path 
$l\simeq 80\:\mu{\rm m}$. Let us assume, following the conclusion of
\cite{umansky97} that the zero-field mobility is determined by
background impurities ({\it i.e.} $l\simeq l_S$), while $l_L\simeq
10l_S$. Using the typical volume concentration of the residual
impurities \cite{saku96}, $n_S^{\rm (3D)}\simeq 2.5\times 10^7\: {\rm
cm}^{-3}$, and the value of the  Bohr radius in GaAs,
$a_B\simeq 10\:{\rm nm}$, we estimate the sheet density of strong
scatterers as $n_S\sim a_B n_S^{\rm (3D)}\simeq (2\:\mu{\rm
m})^{-2}$. With these parameters, we find $B_\tau\simeq 5\:{\rm mT}$,
$B_0\simeq B_{\rm perc}\simeq 60\:{\rm mT}$. The condition of the
diluted array of scatterers, $B_\tau\ll B_{\rm perc}$, is now
well satisfied. Since $B_0\simeq B_{\rm perc}$, the Lorentz-gas
behavior (\ref{e3}) is fully destroyed (the case (C) in our
classification), and the negative MR is determined by
the interplay of smooth disorder and strong scatterers, as described
above. Note that though parametrically $B_0\ll B_{\rm perc}$ for
$l_L\gg l_S$, in practice their values are very close, since 
$B_0/ B_{\rm perc}\simeq (10l_S/l_L)^{1/4}$. The predicted negative
MR may be used for the experimental determination of the
ratio $l_L/l_S$. Strong negative MR has been observed in
very-high-mobility 
samples \cite{umansky97,nmr-private}, in qualitative agreement with
our theory. A more detailed experimental check of our predictions
would be desirable.

In conclusion, we have studied the quasiclassical magnetotransport of a
2DEG with smooth disorder and rare strong
scatterers. Interplay of these two types of disorder leads to a novel
mechanism of strong negative MR; the latter is shown to
saturate with increasing $B$  at a value determined 
solely by smooth disorder. The results are relevant to experiments on
transport in dilute antidot arrays, as well as in high-mobility
heterostructures with
background impurities or interface imperfections.

We thank M.~Heiblum, J.~Smet, V.~Umansky, and D.~Weiss for
discussions of  
experimental results, which motivated this research.
This work was supported by the SFB 195  and the Schwerpunktprogramm
``Quanten-Hall-Systeme'' der Deutschen Forschungsgemeinschaft.

\end{multicols}

\end{document}